\documentclass{jpconf}
\usepackage{amsfonts,amssymb,amstext}
\usepackage{cite}
\usepackage{graphicx}
\usepackage{multirow}

\begin{document}

\sloppy

\title{Augmented kludge waveforms and Gaussian process regression for EMRI data analysis}
\author{Alvin J K Chua}
\address{Institute of Astronomy, University of Cambridge, Madingley Road, Cambridge CB3 0HA, United Kingdom}
\ead{ajkc3@ast.cam.ac.uk}

\begin{abstract}
Extreme-mass-ratio inspirals (EMRIs) will be an important type of astrophysical source for future space-based gravitational-wave detectors. There is a trade-off between accuracy and computational speed for the EMRI waveform templates required in the analysis of data from these detectors. We discuss how the systematic error incurred by using faster templates may be reduced with improved models such as augmented kludge waveforms, and marginalised over with statistical techniques such as Gaussian process regression.
\end{abstract}

\section{Introduction}
The end-2015 launch of the LISA Pathfinder satellite \cite{D2015} is set to usher in a new era of space-based gravitational-wave (GW) detectors such as eLISA \cite{AEA2012} and DECIGO \cite{KEA2011}. Data collected from these detectors will contain a wealth of astrophysical information, which may be used to complement electromagnetic observations \cite{SS2009} and to test strong-field general relativity \cite{GVLB2013}.

One important type of source for space-based detectors is the extreme-mass-ratio inspiral (EMRI) of a stellar-mass compact object (CO) into a massive black hole (BH). The inspiral exhibits strong relativistic effects a few years before the CO plunges; these are imprinted on the GW signal from the source. Matching the signal to a waveform template from an accurate EMRI model will then allow precise estimation of the source's astrophysical parameters.

Accurate EMRI waveforms based on the Teukolsky equation \cite{H2001} and gravitational self-force calculations \cite{B2009} are available, but are extremely expensive to generate. The analysis of detector data will require computationally affordable alternatives such as the analytic kludge (AK) \cite{BC2004} and numerical kludge (NK) \cite{BEA2007} models, which combine mixed prescriptions for computing the orbit and waveform. However, the systematic error in these templates with respect to the true GW signal will lead to reduced detection prospects and inaccurate parameter estimation.

These proceedings are based on a talk given by the author at the 11th Edoardo Amaldi Conference on Gravitational Waves. In this talk, we discussed methods for reducing and marginalising over model error in EMRI waveforms. Sec. \ref{sec:AAK} introduces an augmented kludge waveform that significantly improves on the accuracy of the AK model, with virtually no added computational cost \cite{CG2015}. It uses a frequency map to the three fundamental frequencies associated with a geodesic orbit around a Kerr black hole \cite{S2002}, and features updated evolution equations along with a fit to the slower but more accurate NK model.

While the trade-off between accuracy and computational speed for EMRI waveforms may be lessened with improved models, it cannot be removed completely. However, the statistical technique of Gaussian process regression (GPR) may be used to marginalise over the systematic error in a GW model \cite{MG2014}, which significantly improves parameter estimation performed with faster but less accurate templates \cite{MBCG2016}. The application of GPR to EMRI data analysis is discussed in Sec. \ref{sec:GPR}, where its feasibility is illustrated with a one-parameter likelihood search.

\section{Augmented kludge waveforms}\label{sec:AAK}
Kludge waveform models are semi-relativistic EMRI models designed for robust use in data analysis. The inspiral trajectory in the NK model of Babak et al. \cite{BEA2007} is built from Kerr geodesics and evolved with post-Newtonian (PN) expressions. On orbital timescales, the motion of the CO is approximately geodesic and hence features periapsis precession of the orbital ellipse, along with Lense--Thirring precession of the orbital plane. Both forms of precession are due to discrepancies in the frequencies $\omega_{r,\theta,\phi}$ of the radial, polar and azimuthal components of motion; expressions for these frequencies in terms of the (quasi-Keplerian) orbital parameters $(e,\iota,p)$ of the geodesic have been given by Schmidt \cite{S2002}.

The inspiral trajectory in the AK model of Barack and Cutler \cite{BC2004} is quicker to compute but less accurate, as it consists instead of PN-evolved Keplerian ellipses parametrised by $(e,\iota,p)$. The orbital frequency $f_\mathrm{orb}$ is determined by Kepler's third law, while the two precession rates $f_\mathrm{peri}$ and $f_\mathrm{LT}$ are introduced and evolved separately with additional PN expressions. However, $(f_\mathrm{orb},f_\mathrm{peri},f_\mathrm{LT})$ do not in general equal the correct values $(\omega_r,\omega_\phi-\omega_r,\omega_\phi-\omega_\theta)$. This causes the AK waveform to dephase rapidly relative to the NK waveform, even at the early-inspiral stage.

We propose an augmented AK model \cite{CG2015} that matches the three AK frequencies with the appropriate combinations of $\omega_{r,\theta,\phi}$; this is achieved by mapping the BH mass $M$, the BH spin parameter $a$ and the semi-latus rectum $p$ to some unphysical values $(\tilde{M},\tilde{a},\tilde{p})$, which are given implicitly by the algebraic system of equations
\begin{equation}
(f_\mathrm{orb},f_\mathrm{peri},f_\mathrm{LT})|_{(\tilde{M},\tilde{a},\tilde{p})}=(\omega_r,\omega_\phi-\omega_r,\omega_\phi-\omega_\theta)|_{(M,a,p)}.
\end{equation}
Substituting $(\tilde{M},\tilde{a},\tilde{p})$ for the physical parameters $(M,a,p)$ in the AK model provides an instantaneous correction of the three AK frequencies at any point along the inspiral trajectory.

In the augmented AK model, we evaluate the map at a single trajectory point $(e_0,\iota_0,p_0)$ to obtain the corresponding point $(e_0,\iota_0,\tilde{p}_0)$ on an unphysical trajectory, then evolve the unphysical trajectory as well as the three AK frequencies with higher-order 3PN $\mathcal{O}(e^6)$ expressions given by Sago and Fujita \cite{SF2015}. We also compute two additional points next to $(e_0,\iota_0,p_0)$ on the NK trajectory and evaluate the map at each point, which allows us to perform a local quadratic fit of the unphysical trajectory to the (mapped) NK trajectory for greater accuracy. As a result of this fit, the augmented model features quadratic-in-time evolution of the unphysical BH mass and spin parameters. Since the map-and-fit is only performed locally on the inspiral trajectory, the added computational cost is fixed but essentially insignificant.

\begin{figure}
\centering
\includegraphics[width=0.6\columnwidth]{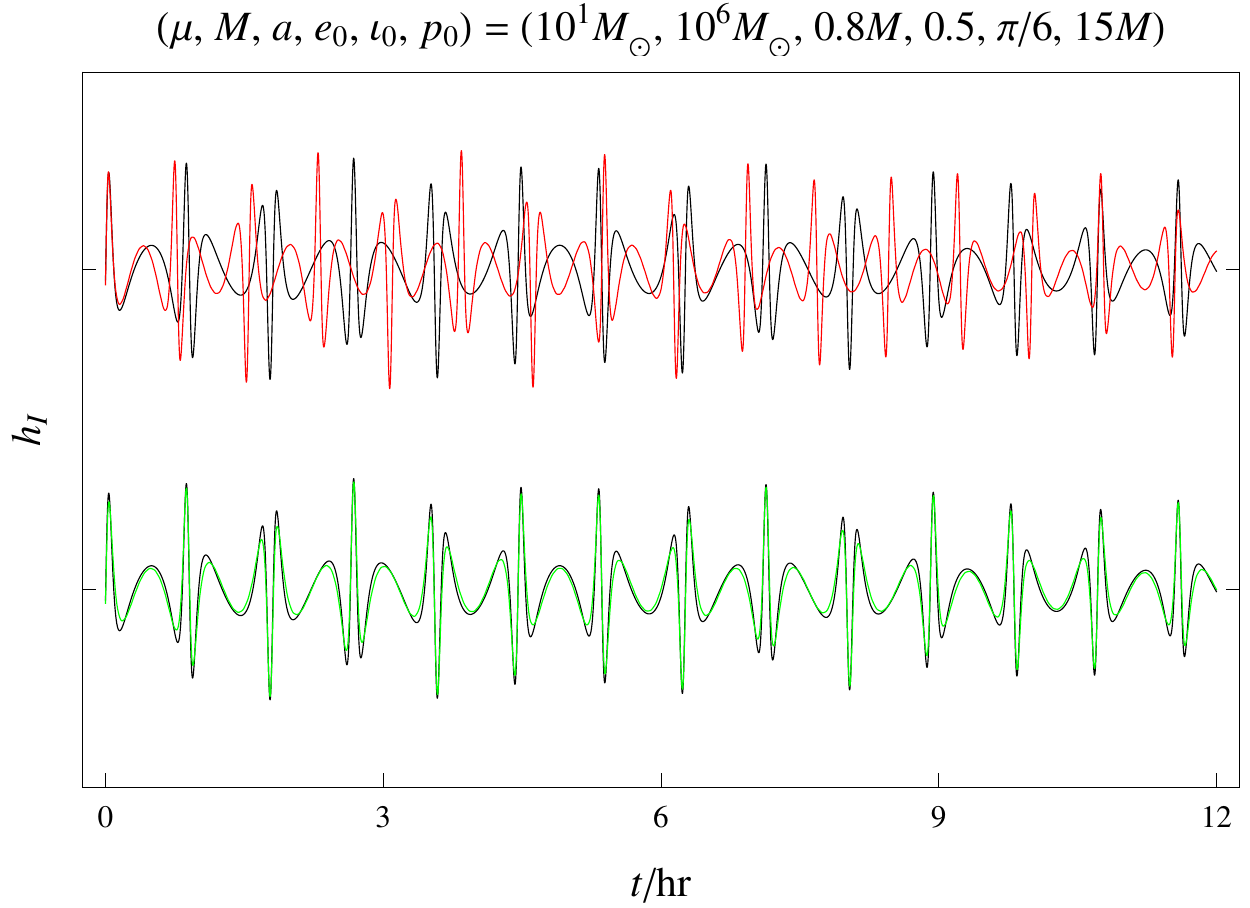}
\caption{First 12 hours of original (red) and augmented (green) AK waveforms overlaid on NK waveform (black), for generic EMRI with initial semi-latus rectum of $15M$.}
\label{fig:waveforms}
\end{figure}

For a generic early-inspiral EMRI with a CO mass of $\mu=10\,M_\odot$, a BH mass of $M=10^6\,M_\odot$, a BH spin of $a=0.8M$, an initial eccentricity of $e_0=0.5$, an initial inclination of $\iota_0=\pi/6$ and an initial semi-latus rectum of $p_0=15M$, the original AK waveform and the NK waveform are a full cycle out of phase within three hours (see top plot of Fig. \ref{fig:waveforms}). The severe dephasing on this timescale is removed by the instantaneous correction of the three AK frequencies in the augmented model (see bottom plot of Fig. \ref{fig:waveforms}). For late-inspiral EMRIs over longer timescales, the improvement in accuracy provided by the augmented AK waveform is just as pronounced. We consider several sources with differing $\mu$, $a$ and $e_0$, choosing $p_0$ for each EMRI such that the CO plunges at $t=1\,\mathrm{yr}$ \cite{CG2015}; the length of time over which the AK waveform remains phase-coherent with the NK waveform is increased from under an hour (in the original model) to over two months (in the augmented model) for all the sources.

\section{Gaussian process regression}\label{sec:GPR}
In GW data analysis, waveform templates $H(\lambda)$ are used to construct the Bayesian likelihood $L(\lambda|x)$ for the model parameters $\lambda$, given the data $x$. However, any theoretical error in the waveform model $H$ will reduce detection signal-to-noise ratios (SNRs) and parameter estimation accuracy. In a recently proposed method for dealing with model error \cite{MG2014}, the likelihood is marginalised over the difference $\delta h$ between the approximate waveform $H$ and an accurate (true) waveform model $h$. This marginalisation is performed analytically by taking the prior for $\delta h$ to be a Gaussian process trained on a small pre-computed set of differences $\{\delta h(\lambda_1),\ldots,\delta h(\lambda_n)\}$. The new GPR likelihood is still cheap to compute (since it depends only on $H$ and the small training set), and will significantly improve parameter estimation prospects for Advanced LIGO and other ground-based GW detectors \cite{MBCG2016}.

Interpolation of the waveform difference $\delta h$ from the training set is central to the GPR method. Hence the performance of the marginalised likelihood depends strongly on the spacing between neighbouring training set points, with the maximal characteristic spacing $\Delta\lambda$ (beyond which the interpolant becomes suboptimal) determined by the accuracy of the approximate model $H$. For GPR to be useful in EMRI data analysis, $\Delta\lambda$ must be significantly larger than the characteristic grid size $\Delta\lambda_\mathrm{TB}$ in a template bank search with the accurate model $h$. In the worst case of an inaccurate approximate model where $\delta h$ has comparable SNR to $H$ and $h$, we have $\Delta\lambda\sim\Delta\lambda_\mathrm{TB}$ for an SNR of 1. However, $\Delta\lambda_\mathrm{TB}$ scales inversely with SNR while $\Delta\lambda$ does not, and so the number of training set points required to estimate the parameters of a typical eLISA source with an SNR of 30 will be $\sim10^9$ times smaller than the number of template bank points required ($\sim10^{30}$ for a fully coherent search in the 14-dimensional parameter space \cite{GEA2004}).

\begin{figure}
\centering
\includegraphics[width=0.6\columnwidth]{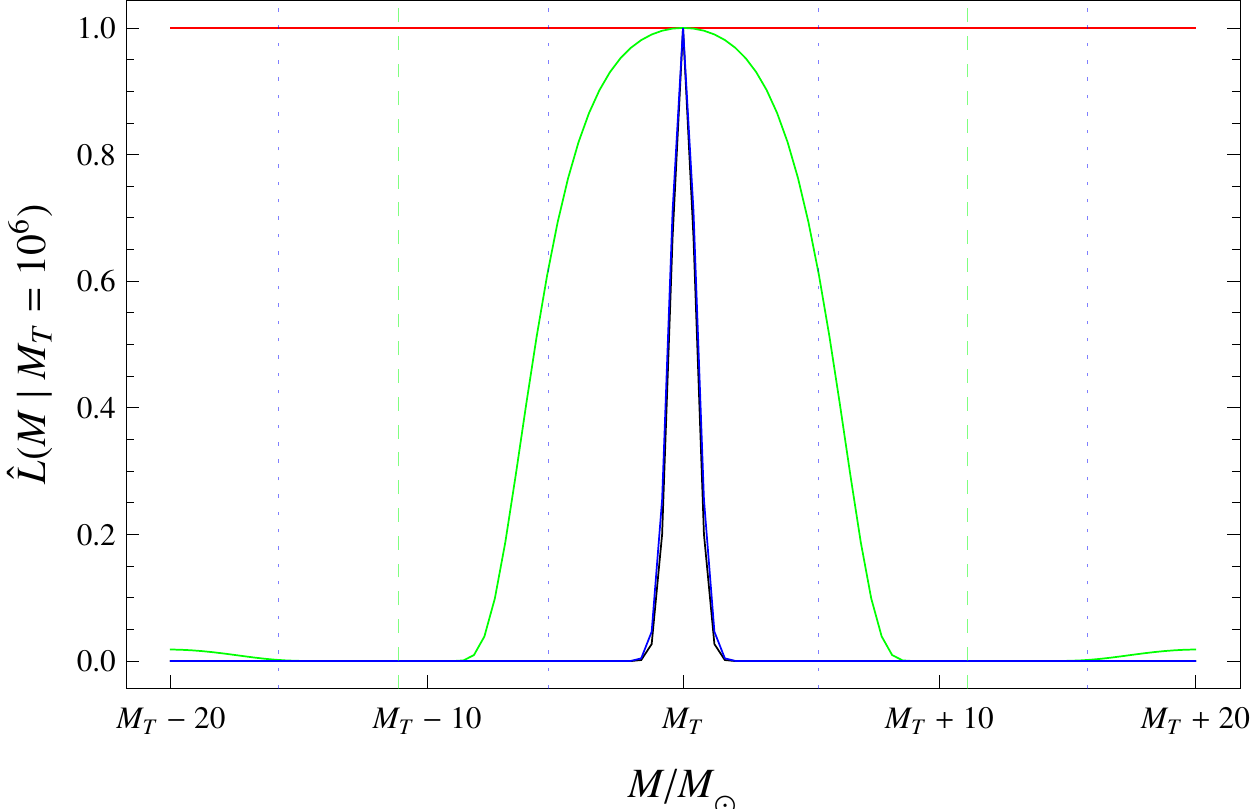}
\caption{Approximate likelihood (red) and marginalised likelihoods with sparse (green) and dense (blue) training sets overlaid on true likelihood (black), for generic EMRI with BH mass of $10^6\,M_\odot$ and SNR of 30. Likelihoods are normalised to peak value of 1. Vertical lines mark positions of points in sparse (dashed) and dense (dotted) training sets.}
\label{fig:likelihoods}
\end{figure}

To illustrate the feasibility of the GPR method for EMRI parameter estimation, we apply it to a one-dimensional search in the BH mass $M$, using the NK and original AK waveforms as accurate and approximate templates respectively. The true signal is taken to be an NK waveform with $M/M_\odot=M_T=10^6$ and an SNR of 30. Fig. \ref{fig:likelihoods} shows the accurate (true) and approximate likelihoods, as well as two marginalised likelihoods with sparse ($\Delta M/M_\odot\approx20$) and dense ($\Delta M/M_\odot\approx10$) training sets. The true likelihood is sharply peaked at the correct value due to the large SNR, while there is no local peak in the approximate likelihood since the original AK model is too inaccurate. Both marginalised likelihoods show a peak at the correct value; the dense training set performs better and gives a near-perfect recovery of the true likelihood, which is far more expensive to sample. Although the two training sets considered in this example are still too dense for likelihood searches over the full parameter space, using improved approximate templates such as the augmented kludge waveforms in Sec. \ref{sec:AAK} should lower the training set density required for the marginalised likelihood to perform optimally.

\section{Conclusion}
We have proposed augmented kludge waveforms and Gaussian process regression as potential methods for reducing and marginalising over error in EMRI waveform models, with a view to improving detection and parameter estimation prospects for future space-based GW detectors.

\section*{Acknowledgements}
This talk was based on work done in collaboration with Jonathan Gair and Christopher Moore at the Institute of Astronomy, Cambridge. The author's work was supported by the Cambridge Commonwealth, European and International Trust.

\section*{References}
\bibliographystyle{unsrt}
\bibliography{references}

\end{document}